\newif\ifarxiv
\arxivtrue
\ifarxiv
\documentclass{emulateapj}
\else
\documentclass[12pt,preprint]{aastex}
\fi
\usepackage{graphicx}
\usepackage{amsmath}
\usepackage{amsfonts}
\usepackage{amssymb}
\usepackage{color}
\usepackage{epsf}

\newcounter{address}
\newcommand{\be}{\begin{equation}}
\newcommand{\ee}{\end{equation}}
\newcommand{\ba}{\begin{eqnarray}}
\newcommand{\ea}{\end{eqnarray}}

\newcommand{\dd}{\mathrm{d}}
\renewcommand{\vec}[1]{{\bf #1}}

\newcommand{\sectionname}{$\mathsection$}
\newcommand{\equationname}{Equation}
\newcommand{\eqnname}{\equationname}

\newcommand{\zacc}{\ensuremath{z_{\mathrm{acc}}}}
\newcommand{\vcrit}{\ensuremath{V_{\mathrm{crit}}}}
\newcommand{\vcirc}{\ensuremath{V_{\mathrm{circ}}}}

\ifarxiv

\fi

\ifarxiv
\submitted{Astrophys.~J., submitted}
\fi

\begin{document}

\title{Low-mass Suppression of the Satellite Luminosity Function Due
  to the Supersonic Baryon--Cold-Dark-Matter Relative Velocity}
\author{Jo Bovy\altaffilmark{1} and Cora Dvorkin}
\affil{Institute for Advanced Study, Einstein Drive, Princeton, NJ 08540, USA}
\altaffiltext{\theaddress}{\stepcounter{address} Hubble fellow}

\begin{abstract}
We study the effect of the supersonic baryon--CDM flow, which has
recently been shown to have a large effect on structure formation
during the dark ages $10 \lesssim z \lesssim 1000$, on the abundance
of luminous, low-mass satellite galaxies around galaxies like the
Milky Way. As the supersonic baryon--CDM flow significantly suppresses
both the number of halos formed and the amount of baryons accreted
onto such halos of masses $10^6 < M_{\mathrm{halo}} / M_\odot < 10^8$
at $z \gtrsim 10$, a large effect results on the stellar luminosity
function before reionization. As halos of these masses are believed to
have very little star formation after reionization due to the effects
of photo-heating by the ultraviolet background, this effect persists to
the present day. We calculate that the number of low-mass $10^6 <
M_{\mathrm{halo}} / M_\odot < 10^8$ halos that host luminous satellite
galaxies today is typically suppressed by 50\,percent, with values
ranging up to 90\,percent in regions where the initial supersonic velocity
is high. We show that this previously-ignored cosmological effect
resolves most of the tension between the observed and predicted number
of low-mass satellites in the Milky Way, obviating the need for any
other mass-dependent star-formation suppression before reionization.
\end{abstract}

\keywords{
        cosmology: theory
        ---
        early universe
        ---
        galaxies: formation
        ---
        Galaxy: halo
        ---
        galaxies: statistics
        ---
        Galaxy: structure
}

\section{Introduction}

A robust prediction of the Cold Dark Matter (CDM) cosmological paradigm of hierarchical
clustering is that the halos of galaxies like the Milky Way should
contain hundreds of satellite subhalos that could be expected to host
observable galaxies \citep{Kauffmann93a,Klypin99a,Moore99a}. The
success of the $\Lambda$CDM model from the horizon-scale
\citep{Komatsu11a} to the small scale of the Lyman-$\alpha$ forest
\citep{Viel08a} suggests that this ``missing satellites problem'' is
most likely the result of the details of galaxy-formation physics in
low-mass halos rather than a manifestation of a deviation from the
standard framework on small scales.

Potential solutions to the missing satellites problem have so far come
in two flavors. One approach aims to reduce the intrinsic small-scale
power in the CDM framework by positing alternatives to standard
inflationary mechanisms for producing the initial perturbation
spectrum \citep{Kamionkowski00a}. Similarly, warm dark matter
models naturally suppress the number of small, bound structures in the
Universe \citep{Polisensky:2010rw}. The alternative to modifying the
cosmological framework is to invoke astrophysical explanations for
suppressing star formation in low-mass galaxies. In particular,
suppression of star formation by photo-heating by the ultraviolet (UV)
background after reionization naturally leads to a
star-formation-efficiency cut-off at approximately the correct mass
scale \citep{Bullock00a,Somerville02a}. This can explain
the number counts of ``classical dwarf spheroidals'' (dSphs) around
the Milky Way ($M_V \lesssim -6$; \citealt{Koposov09a}).

The Sloan Digital Sky Survey (SDSS; \citealt{2000AJ....120.1579Y})
discovered many more smaller dSphs in the Milky Way's halo
\citep[e.g.,][]{2005ApJ...626L..85W,2006ApJ...647L.111B,2006ApJ...650L..41Z,
  2007ApJ...654..897B,2007ApJ...656L..13I,2007ApJ...662L..83W}. To
explain the number counts of these new discoveries within
$\Lambda$CDM, it is necessary to invoke additional suppression of star
formation in the lowest mass halos \emph{before} reionization
\citep[e.g.,][]{Koposov09a}, as star formation in these halos mostly
ends at reionization due to photoionization. Specifically, the stellar
mass in low-mass halos ($M \lesssim 10^8 M_\odot$) can only be
$\lesssim$ 10$^{-4}$ times the halo mass \citep{Madau08a}, an order of
magnitude less than the stellar-mass fraction in higher mass
halos. This is only $\approx 0.05$\,\% of the universal baryon
fraction, while typically we would expect star formation efficiencies
of at least a few percent, even when star formation only proceeds
through $H_2$ cooling \citep{Bovill09a,Salvadori09a} and it is unclear
whether radiative feedback from the first generation of stars
suppresses pre-reionization star formation in low-mass halos or not
\citep{1996ApJ...467..522H,1999ApJ...518...64O,2000ApJ...534...11H,2002ApJ...575...33R,2002ApJ...575...49R,2008ApJ...679..925W}.

In this paper we determine the influence of the previously-ignored
effect of the relative velocity between baryons and dark matter at
recombination on the stellar content of the smallest-mass galaxies. As
was recently pointed out by \citet{TH10a}, the typical baryon--CDM
relative velocity of 30 km s$^{-1}$ at recombination is supersonic as
the baryon sound speed at kinetic decoupling drops to 5 km
s$^{-1}$. While the relative velocity decays as $\propto a^{-1}$,
where $a$ is the scale factor, the supersonic baryon flow has a
profound effect on the formation of the first structures
\citep{TBH,2011ApJ...730L...1S,2011MNRAS.412L..40M,2011ApJ...736..147G,Naoz12a,2012arXiv1201.1005V,Oleary12a},
which through the effects of hierarchical clustering might persist
today (e.g., in the baryon acoustic feature, \citealt{Dalal10a}), and it
could also give rise to CMB B-modes if the effect persists during the
epoch of reionization \citep{Ferraro:2011nc}.  We show that the
supersonic baryon flow has a large effect on both the number and the
baryon content of halos with masses between the $H_2$ cooling limit
($\approx 10^6 M_\odot$; \citealt{1997ApJ...474....1T}) and the
photo-ionization scale ($\approx 10^7 M_\odot$ at $z = 11$,
\citealt{2000ApJ...542..535G}). This typically reduces the number of
satellites by 50 up to 90\,percent at $M_V \approx -1$.

The outline of this paper is as follows. In \sectionname~\ref{sec:vbc}
we review the supersonic baryon--CDM relative velocity effect. We then
compute the pre-reionization mass and luminosity functions in
\sectionname~\ref{sec:lumfunc}. We discuss the low-redshift behavior
of the satellite luminosity function in
\sectionname~\ref{sec:lowz}. Our conclusions are in
\sectionname~\ref{sec:conclusion}. We assume cosmological parameters
matching the WMAP 7-year data \citep{Komatsu11a}: $\Omega_{b,0}=
0.0456$, $\Omega_{c,0} = 0.227$, $z_{\mathrm{eq}} = 3232$, $H_0= 70.4$
km s$^{-1}$ Mpc$^{-1}$, $\Delta^2\zeta (k = 0.002\ {\rm Mpc}^{-1}) =
2.44 \times10^{-9}$, and $n_s = 0.963$.

\section{Linear evolution of density perturbations in the presence of a supersonic flow}\label{sec:vbc}

After kinetic decoupling ($z \approx 1020$; \citealt{Eisenstein98a})
the baryonic sound speed drops to 5 km s$^{-1}$, while the baryons
move relative to the CDM with a typical velocity of 30 km s$^{-1}$. As
originally pointed out by \citet{TH10a}, this means that second-order
terms such as $\vec{v}\cdot\nabla\delta$ and
$\vec{v}\cdot\nabla\vec{v}$, which are typically neglected to first
order, become as large as first-order terms in the continuity and
Navier-Stokes equations that describe the evolution of inhomogeneities
in the baryons and CDM after recombination. We follow the treatment of
\citet{TBH}, who shows that the supersonic relative flow
$\vec{v}_{bc}$ is homogeneous over a few comoving Mpc, with
$\vec{v}_{bc}$ drawn from a Gaussian with a variance per axis of
$\sigma_{bc}^2/3$ with $\sigma_{bc} = 30$ km s$^{-1}$ at kinetic
decoupling. The evolution equations for the relative CDM ($\delta_c$)
and the baryon ($\delta_b$) perturbations can be written in the baryon
rest frame in a $\sim$ comoving Mpc patch with a supersonic flow
$\vec{v}_{bc}$ as

\ba\label{eq:evolution}
{\partial\delta_c \over \partial t} &=& {i\over a}{\bf v}_{bc}.{\bf k}\delta_c-\theta_c \nonumber\\
{\partial\theta_c \over \partial t} &=& {i \over a}{\bf v}_{bc}.{\bf k}\theta_c - {3H_0^2\over 2}{\Omega_{m,0} \over a^3}({\Omega_{c,0}\over \Omega_{m,0}}\delta_c + {\Omega_{b,0} \over \Omega_{m,0}}\delta_b)\nonumber\\
&&\quad -2H\theta_c \nonumber\\
{\partial\delta_b \over \partial t} &=& -\theta_b \nonumber\\
{\partial\theta_b \over \partial t} &=& -{3H_0^2\over 2}{\Omega_{m,0} \over a^3}({\Omega_{c,0} \over \Omega_{m,0}}\delta_c+{\Omega_{b,0} \over \Omega_{m,0}}\delta_b)\nonumber\\
&& \quad -2H\theta_b+{k^2\over a^2}{k_B T_b \over \mu m_H}(\delta_b + \delta_{T_b})\,,
\ea

where we write the baryonic sound speed as $c_s^2=k_B T_b(\delta_b +
\delta_{T_b})/\mu m_H \delta_b$, $\theta$ is the velocity divergence,
$\Omega_{m,0}=\Omega_{b,0}+\Omega_{c,0}$, $\mu$ = 1.22 is the mean
molecular weight including a helium mass fraction of 0.24, $m_H$ is
the mass of the hydrogen atom, and $T_b$ and $\delta_{T_b}$ are the
mean baryon temperature and its relative fluctuation,
respectively. The evolution of $T_b$ is given by \citep{TH10a}

\be
T_b={T_{\rm CMB,0}\over a}\left[1+{a/(1/119)\over 1+(1/115/a)^{3/2}}\right]^{-1}\,,
\ee

where $T_{\rm CMB,0} = 2.726$ K. The temperature fluctuation evolves
according to \citep{Barkana05a,Naoz05a}

\be\label{eq:deltaT}
{\partial \delta_{T_b} \over \partial t} = -{2 \over 3}\theta_b  - {x_e(t) \over t_\gamma} {1 \over a^4}{T_\gamma \over T_b} \delta_{T_b}\,,
\ee

where $x_e(t)$ is the free electron fraction out of the total number
density of gas particles, $t_\gamma = 8.55 \times 10^{-13}$ yr$^{-1}$,
$T_\gamma$ is the mean photon temperature, and we have neglected the
photon inhomogeneities compared to \citet{Naoz05a}. We obtain the free
electron fraction $x_e(t)$ using RECFAST \citep{Seager99a,Seager00a}.

We solve the complex system of \eqnname
s~(\ref{eq:evolution}--\ref{eq:deltaT}) by taking initial conditions
for the matter inhomogeneities and velocities from
CAMB\footnote{\url{http://camb.info/}~.} \citep{camb} at kinetic
decoupling ($z = 1020$) and setting up the initial inhomogeneities in
the baryon temperature following the approximation done in
\citet{Naoz05a} by requiring $\partial \delta_{T_b} / \partial t =
\partial \delta_{T_\gamma} / \partial t$, where $\delta_{T_\gamma}$ is
the relative photon temperature perturbation. This approximation is
justified because of tight thermal coupling between the baryons and
the photons and it has been shown to affect the power spectra only by
a fraction of a percent at $z=1020$. The relative velocity in
\eqnname~(\ref{eq:evolution}) decays as $\vec{v}_{bc} \propto a^{-1}$.

\section{Luminosity function before reionization}\label{sec:lumfunc}

We follow the procedure of \citet{TBH} to calculate the halo mass
function and the baryonic content of low-mass halos before
reionization in the presence of a supersonic baryon--CDM flow. We
calculate the halo mass function using the Extended Press-Schechter
formalism \citep{Bond:1990iw,Bower:1991kf,1993MNRAS.262..627L}

\be 
\frac{\dd N }{\dd
  M_{\mathrm{halo}}}(M_{\mathrm{halo}};v_{bc}) =
\frac{\bar{\rho}_0}{M_{\mathrm{halo}}}\,\left|\frac{\dd S}{\dd
  M_{\mathrm{halo}}}\right|\,f(\delta_c(z),S)\,, 
\ee 
where $\bar{\rho}_0$ is the mean matter density in the universe and $S$ is the usual variance given by 

\be 
S(M;v_{bc}) = \int d\ln k\Delta_m^2(k;v_{bc})|W(k;R)|^2
\ee

\ifarxiv
\begin{figure}[h!]
\includegraphics[width=0.5\textwidth,clip=]{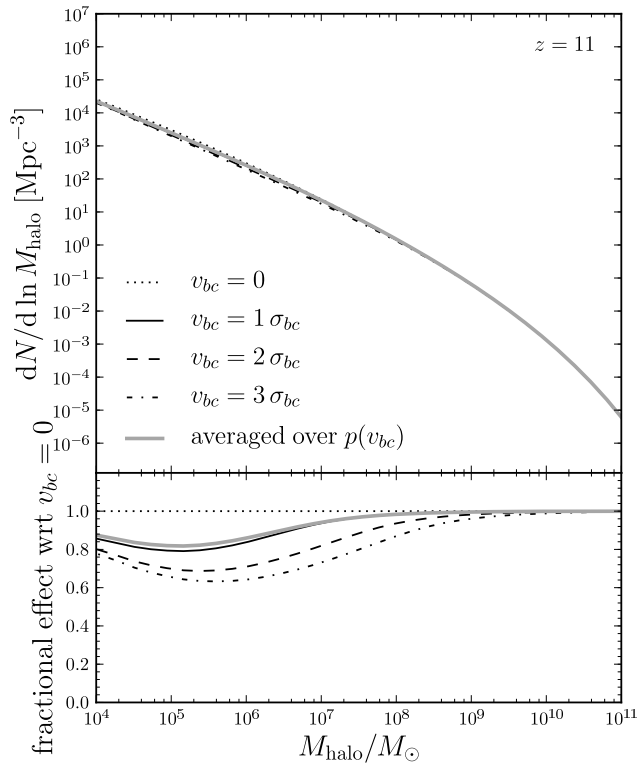}
\caption{Halo mass function at $z = 11$ with a relative baryon-CDM
  velocity of $v_{bc} = 0$ and $v_{bc} \neq 0$. 1 $\sigma_{bc}$
  corresponds to approximately 30 km s$^{-1}$ at decoupling ($z =
  1020$). The globally-averaged effect obtained by averaging over the
  Gaussian probability distribution function of $\vec{v}_{bc}$ is
  shown as the gray curve.}\label{fig:z11mass}
\end{figure}

Here $\Delta_m^2(k;v_{bc})$ is the matter power spectrum computed
using \eqnname s~(\ref{eq:evolution}--\ref{eq:deltaT}).  The transfer
functions computed using \eqnname
s~(\ref{eq:evolution}--\ref{eq:deltaT}) depend on the angle between
$\vec{k}$ and $\vec{v}_{bc}$. In what follows, we average all
solutions for $|\delta_b|$ and $|\delta_c|$ over this angle and give
results for $v_{bc} = | \vec{v}_{bc}|$.  $W(k;R)$ is the tophat window
function and the initial comoving radius $R$ corresponds to a halo of
mass $M_{\mathrm{halo}} = 4\pi R^3\bar{\rho}_0/3$. We use the
functional form from \citet{Sheth99a} for $f(\delta_c(z),S)$: \be
f(\delta_c(z),S) =
A\frac{\nu}{S}\sqrt{\frac{a}{2\pi}}\left[1+\frac{1}{(a\nu^2)^q}\right]\,e^{-a\,\nu^2/2}\,,
\ee where $\nu = \delta_c(z)/\sqrt{S}$, $a = 0.75$, $q = 0.3$, $A =
0.322$ \citep{Sheth02a}, and $\delta_c(z)=1.67$ is the critical
density of collapse at $z \approx 10$.

\begin{figure}[h!]
\includegraphics[width=0.5\textwidth,clip=]{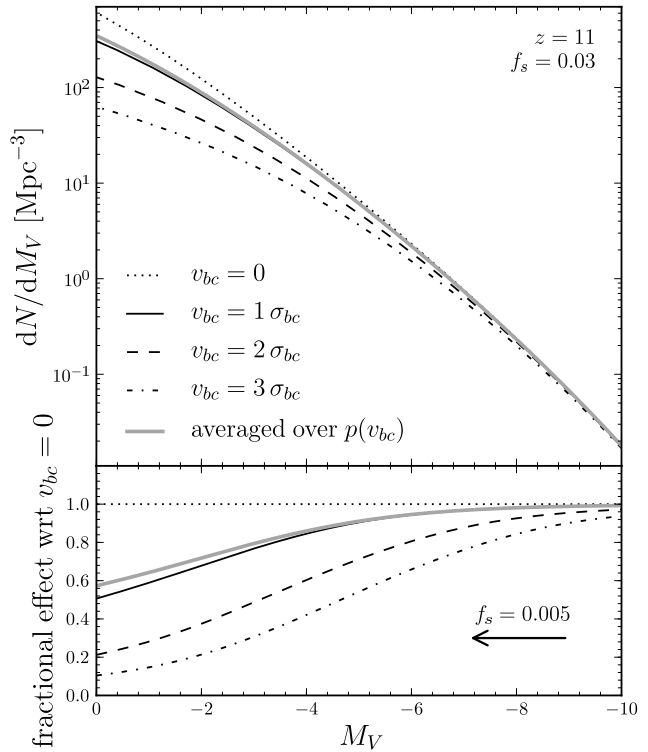}
\caption{Luminosity function at $z = 11$ using the luminosity that the
  halos would have today. This is the total effect on the luminosity
  function from the combination of the suppression of the halo mass
  function and the reduction in the gas fraction due to $v_{bc} \neq
  0$. The average case is shown in the gray curve. The separate contributions to the suppression for $v_{bc} \neq
  0$ from the halo mass function and gas fraction are shown in
  \figurename~\ref{fig:z11lumfunc-comps}. The arrow in the bottom
  panel shows how far each curve moves when lowering the
  star-formation efficiency to $f_s = 0.005$.}\label{fig:z11lumfunc}
\end{figure}
\fi

Constraints from the Cosmic Microwave Background on the epoch of
reionization are consistent with a redshift of reionization of $z =
11$ \citep{Komatsu11a,Zahn11a} (defined as the redshift at which
reionization would begin if the universe was reionized
instantaneously). The halo mass function at $z = 11$ calculated using
the procedure given above is shown in
\figurename~\ref{fig:z11mass}. This figure shows that the effect of
the supersonic baryon--CDM velocity typically suppresses the number of
dark matter halos by 10 to 20\,percent in the range $10^5 <
M_{\mathrm{halo}} / M_\odot < 10^7$ at $z=11$. $N$-body simulations by
\citet{Naoz12a} show a similar level of suppression in the mass
function at $z=11$.

To estimate the stellar mass in low-mass subhalos at reionization, we
use the results from \citet{Naoz07a} and \citet{Naoz09a}. These
studies have shown that the gas fraction in halos at high-redshift is
suppressed with respect to the universal baryon fraction by the
combined effect of the remaining suppression after recombination in
the baryon density perturbations on small scales
\citep{Naoz07a,Barkana11a} and the baryonic pressure, which gives rise
to a redshift-dependent `filtering' scale
\citep{Gnedin98a,2000ApJ...542..535G,Naoz07a}. This filtering scale
corresponds to the length scales below which the
baryon--to--total-matter fluctuation drops substantially below its
large-scale value, and simulations find that it characterizes the
minimum halo mass that can retain its gas
\citep{Naoz09a,Oleary12a}. To determine the filtering scale $k_F$ as a
function of the supersonic flow velocity $\vec{v}_{bc}$ we fit the
functional form \be \frac{|\delta_b|}{|\delta_{\mathrm{tot}}|} =
(1+r_{\mathrm{LSS}})\left(1+\frac{1}{n}\frac{(k^2/k_F^2)}{1+r_{\mathrm{LSS}}}\right)^{-n}\,,
\ee where $|\delta_{\mathrm{tot}}| = (\Omega_{b,0}\,|\delta_b| +
\Omega_{c,0}\,|\delta_c|) / \Omega_{m,0}$, $r_{\mathrm{LSS}}$ is
determined from the behavior of $|\delta_b|/|\delta_{\mathrm{tot}}|$
on larger scales, $1 \leq k\ \mathrm{Mpc} \leq 10$, and $k_F$ and $n$
are subsequently fit at $k \geq 1$ Mpc$^{-1}$.

Using the filtering scale $k_F$, the filtering mass is defined as \citep{Naoz07a}
\be
M_F(M_{\mathrm{halo}};v_{bc}) = \frac{4\,\pi}{3}\bar{\rho}_0\,\left(\frac{\pi}{k_F}\right)^3
\ee

\ifarxiv
\begin{figure}[th]
\includegraphics[width=0.5\textwidth]{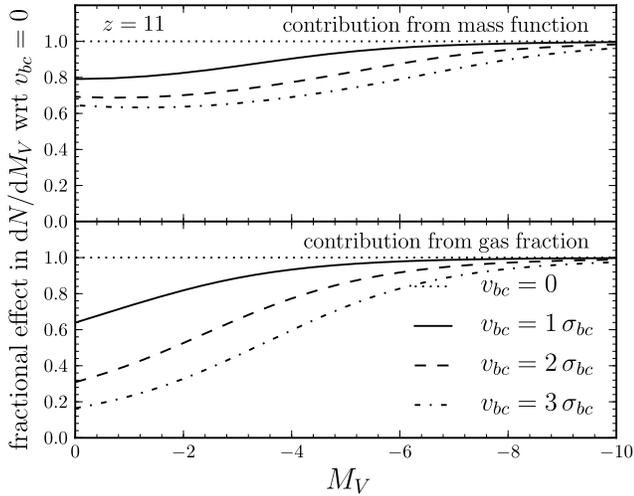}
\caption{Separate contributions to the difference in the $z=11$
  luminosity function in \figurename~\ref{fig:z11lumfunc} between
  $v_{bc} = 0$ and $v_{bc}\neq 0$ from the suppression in the halo
  mass function in \figurename~\ref{fig:z11mass} and the suppression
  in the gas fraction.}\label{fig:z11lumfunc-comps}
\end{figure}
\fi

Using the filtering mass, we can estimate the gas fraction at
redshifts $z \geq 11$ as \citep{Naoz09a} \be\label{eq:fgas}
f_{\mathrm{gas}}(M;v_{bc}) = f_{b,0} \left[ 1 + (2^{\alpha/3} -
  1)\,\left(\frac{M_F}{M}\right)^\alpha\right]^{-3/\alpha}\, \ee where
$f_{b,0}$ is the gas fraction in the high-mass limit. $f_{b,0}$ is
typically lower than the universal baryon fraction because density
perturbations in the baryons at high redshift remain suppressed due to
the lingering effect of the coupling between baryons and photons
before recombination. In \citet{Barkana11a}, the authors show that
$f_{b,0} \approx (1 + 3.2 r_{\mathrm{LSS}}) \Omega_{b,0} /
\Omega_{m,0}$. Following \citet{TBH}, we use $\alpha = 0.7$, even
though this value of $\alpha$ was calculated for halos at $z \approx
20$. As our main objective is to show the difference between the
$v_{bc} = 0$ and $v_{bc} \neq 0$ cases, the exact functional form in
\eqnname~(\ref{eq:fgas}) does not matter greatly.

We assume that a fraction $f_s$ of the gas in a halo---independent of halo mass---turns into stars, such that the stellar mass is given by
\be\label{eq:mstar}
M_s(M_{\mathrm{halo}};v_{bc}) = f_s\,f_{\mathrm{gas}}(M_{\mathrm{halo}};v_{bc})\,M_{\mathrm{halo}}\,.
\ee
We convert this stellar mass into a luminosity by assuming that this stellar mass is turned into stars with very low metallicity $Z = Z_\odot / 200$. Today, such a population would shine with an absolute magnitude $M_V = 6.7$ per solar mass \citep{Madau08a,Bruzual03a} and it is this luminosity that we use to show the resulting luminosity function. 

\figurename~\ref{fig:z11lumfunc} shows the luminosity function at $z =
11$ using the luminosity that the halos would have today. That is, it
is the luminosity function we would observe today if the low-mass
halos were frozen in their pre-reionization state and there was no
evolution in the number of halos. We will study their evolution
in \sectionname~\ref{sec:lowz}. 

\figurename~\ref{fig:z11lumfunc}
assumes a star formation efficiency of $f_s = 0.03$, which gives a
total-to-stellar mass at the high-mass end of 10$^{-2.8}$. The arrow
in the bottom panel of \figurename~\ref{fig:z11lumfunc} shows the
effect of lowering the star-formation efficiency to $f_s = 0.005$. In
this case, halos of a given luminosity come from higher total-mass
halos, where the effect of the supersonic baryon--CDM velocity is
smaller, thus shifting the effect toward smaller luminosities.

\figurename~\ref{fig:z11lumfunc-comps} shows the relative contribution
to the total effect in \figurename~\ref{fig:z11lumfunc} from the
suppression of the mass function in the presence of the supersonic
baryon--CDM flow (see \figurename~\ref{fig:z11mass}) and the
suppression of the baryon fractions. It is interesting to note that
the suppression in the accretion of baryons onto dark matter halos is
the main driver of the suppression of low-luminosity halos.

\citet{2012arXiv1204.1345M} find that shock heating can raise the
baryon temperature by approximately 10\,percent compared to the
evolution that we assume. If we approximate the evolution in $T_b$
that they find by raising $T_b$ by 10\,percent at $z < 20$, we find a
negligible influence on the matter power spectrum and filtering mass
at $z = 11$, such that the conclusions of this section are unaffected
by shock heating.

\section{Present-day satellite luminosity function}\label{sec:lowz}

In this section we quantify how the suppression at the faint end of
the luminosity function before reionization affects the present-day
subhalo luminosity function for a Milky-Way sized halo. We estimate
the satellite luminosity function by running merger-tree simulations
using the Extended Press-Schechter formalism of
\citet{1993MNRAS.262..627L} and \citet{2000MNRAS.319..168C} for a
parent halo of mass $M = 10^{12}\ M_\odot$
\citep{2008ApJ...684.1143X}. We create 100 merger trees for each of
$v_{bc} = 0, 1, 2$, and 3 $\sigma_{bc}$ by computing the $z=0$
$S(M;v_{bc})$ as before and using the linear overdensity for collapse
$\delta_c = 1.67$, extrapolated to $z=0$ using the growth factor
\citep{1996MNRAS.282..263E,1992ARA&A..30..499C}. We resolve subhalos
down to a resolution mass of $M_{\mathrm{res}} = 10^6 M_\odot$ and run
the simulation until $z=11$. As we are only interested in low-mass
subhalos, we only track the mass evolution of subhalos, without
resolving it into sub-subhalos. For each subhalo, we record the
pre-reionization mass $M(z=11)$, defined as the mass of the subhalo at
$z=11$, and the mass at the time of the merger with the parent halo
$M(z=\zacc)$.

\ifarxiv
\begin{figure}[h!]
\includegraphics[width=0.5\textwidth,clip=]{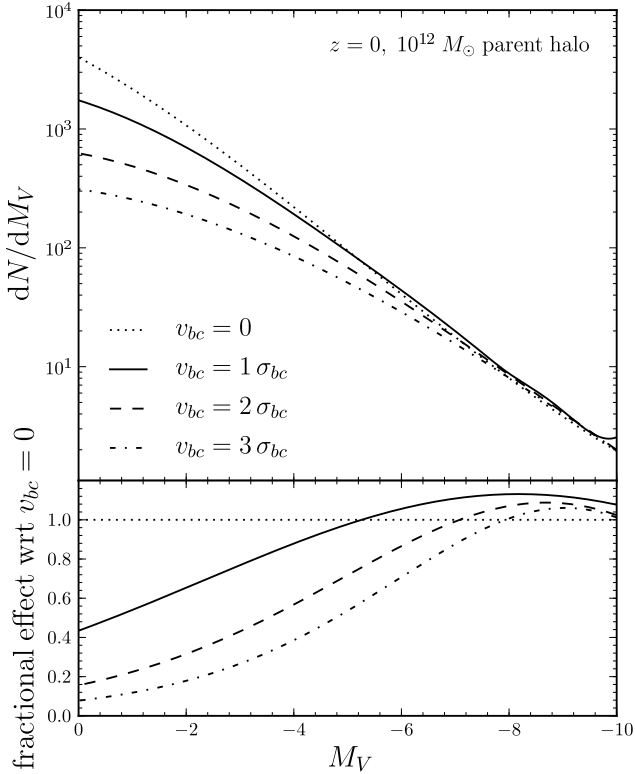}
\caption{Satellite luminosity function at $z = 0$ for a Milky-Way size
  halo ($M = 10^{12}\,M_\odot$). The curves show the average
  luminosity function of 100 merger trees for each value of
  $v_{bc}$. The bottom panel shows the fractional effect with respect
  to $v_{bc} = 0$.}\label{fig:z0lumfunc}
\end{figure}
\fi

We do not follow the dynamical and mass evolution of the subhalos
after they merge with the parent halo. This approximation assumes that
low-mass ($<10^8 M_\odot$) satellites do not get tidally disrupted and
do not lose a significant amount of stellar mass after merging with
the parent halo. High-resolution $N$-body simulations find that only a
few percent of $M > 10^7 M_\odot$ subhalos are tidally destroyed
between $z=1$ and $z=0$ and that subhalos with $M < 10^8 M_\odot$
retain most of their total mass \citep{2007ApJ...667..859D}. Even when
the outer parts of the subhalos are tidally stripped, the stars and
the inner part of the subhalo are stripped only at the last stage of
the tidal disruption of the subhalo \citep{2008ApJ...673..226P}, such
that most of the stellar mass is retained, even if a significant part
of the dark matter halo is tidally stripped. Therefore, to a good
approximation, the stellar-mass function of low-mass satellites should
not be strongly affected by the effects of tidal stripping and
disruption. 

We compute the luminosity of each subhalo by using the prescription of
\eqnname~(\ref{eq:mstar}) applied to the pre-reionization mass
$M(z=11)$ to calculate the pre-reionization stellar-mass and turning
this into a luminosity again using $M_V = 6.7$ per solar mass. We
calculate the post-reionization stellar mass for each subhalo by using
a star-formation efficiency that takes into account suppression by
photoionization
\citep{2000ApJ...542..535G,2006MNRAS.371..401H,2008MNRAS.390..920O}
and that assumes that the effect of the supersonic baryon--CDM
velocity has no influence on star formation after reionization:

\be
M_s(z < 11) = f_*\frac{(M(z=\zacc) - M(z=11))}{\left(1+0.26\left(\vcrit/\vcirc(\zacc)\right)^3\right)^3}\,,
\ee

where $f_* = 10^{-3}/6.25$ and $\vcrit = 35$ km s$^{-1}$ (as in
\citealt{Koposov09a}). We calculate $\vcirc(\zacc)$ by using the
virial radius of Equation~(1) in \citet{Koposov09a}. The
post-reionization contribution to the total stellar luminosity of a
satellite is then calculated by assuming a solar mass-to-light ratio
\citep{2008ApJ...684.1075M}.

\ifarxiv
\begin{figure}[h!]
\includegraphics[width=0.5\textwidth,clip=]{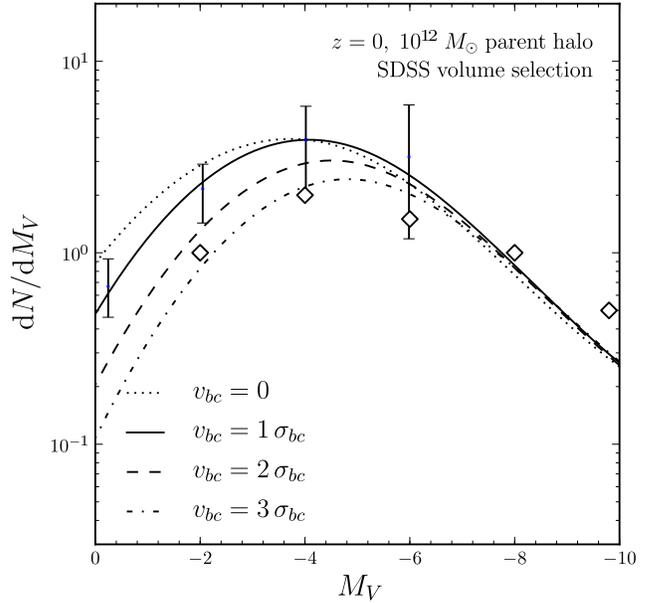}
\caption{Luminosity function of satellite galaxies of the Milky Way
  observable by the SDSS. The diamond data points are taken from
  \citet{Koposov09a}. The error bars on the $v_{bc} =
  1\,\sigma_{bc}$ model curve show the 68\,percent spread in the
  merger trees at luminosities where the predictions with different
  $v_{bc}$ differ; this spread is similar for all four model
  curves. The error bars are highly correlated; for example, the
  correlation between the $M_V = -4$ and $M_V = -2$ error bars is
  0.75. For this figure we assume $f_{\mathrm{s}} =
  0.01$.}\label{fig:sdsslumfunc}
\end{figure}
\fi

The luminosity function of satellites of a Milky-Way type halo
computed in this way is shown in \figurename~\ref{fig:z0lumfunc}. A
comparison between this figure and \figurename~\ref{fig:z11lumfunc}
confirms that the pre-reionization effect on the suppression of the
luminosity function at $z=11$ translates into almost the same
suppression in the satellite luminosity function at $z=0$. Additional
suppression of the luminosity function results from the fact that in a
region of a highly supersonic baryon--CDM flow, structure formation is
delayed, such that a larger fraction of the present-day mass of a halo
was accreted after the epoch of reionization, where it is affected by
the suppressed gas accretion and cooling in the presence of the
photoionizing background after reionization. This additional effect
also explains the slight increase in the number of brighter satellites
in \figurename~\ref{fig:z0lumfunc}. At the bright end---where
pre-reionization suppression due to the supersonic baryon--CDM flow is
small---a larger fraction of a satellite's mass is accreted after
reioinization where we have assumed a mass-to-light ratio appropriate
for a metal-poor population rather than that of an extremely
metal-poor population, which we assumed for pre-reionization star
formation. This discontinuity, which is an artifact of the simplicity
of our assumptions, stretches the $v_{bc} \neq 0$ curves toward the
brighter end, leading to an increased number of bright satellites.

To compare the predicted satellite luminosity function to the observed
luminosity function of satellites of the Milky Way (e.g.,
\citealt{2008ApJ...686..279K}), we assume that the spatial
distribution of the satellites follows a Navarro-Frenk-White profile
with a scale radius of 30 kpc; in accordance with numerical
simulations \citep{2007ApJ...667..859D}, we anti-bias this
distribution radially by multiplying by the Galactocentric radius. We
then calculate the observational fraction by using the simple model
for the SDSS satellite selection function of \citet{Koposov09a}, where
a satellite of absolute magnitude $M_V$ can be detected out to a
distance of $D_{\mathrm{max}} = 10^{1.1-0.228\,M_V}$ kpc from the Sun
(assumed to be 8 kpc from the Galactic center,
\citealt{2009ApJ...704.1704B}), integrating out to a virial radius of
260 kpc. We multiply this selection fraction by 0.194 to account for
the partial sky coverage of the SDSS. All satellites down to $M_V
\approx -6$ could be observed throughout the virial volume by the
SDSS.

The luminosity function of satellites detectable by the SDSS thus
computed is shown in \figurename~\ref{fig:sdsslumfunc}. Even though,
for this figure we have assumed a star-formation efficiency
$f_{\mathrm{s}}$ of only 0.01, it is clear that the supersonic
baryon--CDM velocity has a large effect on the faint end of this
luminosity function that significantly lowers the predicted fraction
of satellites at $M_V > -6$. Thus, the inefficient gas accretion at
high redshift induced by a large supersonic flow has a large and
lasting effect on the abundance of luminous satellites for a galaxy
like the Milky Way.

We do not show the upper limit of $\dd N / \dd M_V < 0.1$ at $M_V = 0$
from \citet{Koposov09a} in \figurename~\ref{fig:sdsslumfunc}, as the
predictions at $M_V = 0$ strongly depend on the exact form of the SDSS
selection function. At $M_V = 0$, the maximum distance out to which a
satellite can be detected by the SDSS is approximately 12 kpc. As we
do not follow the dynamical evolution of the satellites in the parent
halo, our simulations do not capture the fact that those satellites
that come within 20 kpc of the Galactic center have a much larger
chance of being tidally disrupted, such that our predictions at $M_V =
0$ over-estimate the number of observable satellites.

\section{Conclusion}\label{sec:conclusion}

As pointed out by \citet{TH10a}, at recombination, baryons move with a
typical supersonic velocity of 30 km s$^{-1}$ with respect to the dark
matter. While this relative baryon--CDM velocity decays as $\propto
a^{-1}$, it has a large effect on the formation of the first
structures in the Universe. We have investigated the effect on the
present-day abundance of luminous low-mass satellite galaxies
($\lesssim 10^8 M_\odot$) of a Milky-Way size galaxy. One would expect
the supersonic baryon--CDM velocity to affect the abundance and
luminosity of low-mass satellites, as these satellites must form many
of their stars before reionization because photo-heating by the UV
background after reionization suppresses star formation in halos with
circular velocities $\lesssim 30$ km s$^{-1}$. The effect of the
supersonic baryon--CDM flow should be a standard cosmological effect
that needs to be taken into account when predicting the number of
luminous satellite galaxies---the only unknown being the initial value
of the relative velocity in the patch of the Universe that is being
considered. Additionally, we are motivated by the apparent discrepancy
between the observed number of dwarf satellite galaxies in the Milky
Way at $M_V \gtrsim -4$, which indicates that to fully explain the
missing satellites problem, star formation \emph{before} reionization
needs to be suppressed, in addition to the---now
standard---suppression by photo-heating after reionization.

We have shown by following the linear evolution of density
fluctuations while taking into account the non-linear effect of the
supersonic baryon--CDM flow, that the stellar-mass function before
reionization ($z \approx 11$) is significantly suppressed at the
low-luminosity end, with a typical suppression by 50\,percent for
luminosities today of $M_V \gtrsim -4$, going up to 90\,percent for
regions with a high initial supersonic relative velocity. This
suppression is mainly the result of the reduced accretion of gas onto
low-mass halos due to the supersonic flow, with a small contribution
from the overall suppression of the halo mass function at $10^5 <
M_{\mathrm{halo}} / M_\odot < 10^7$. While these calculations use the
linear evolution equations for perturbations, cosmological
high-resolution hydrodynamical simulations of the formation of the
first structures have shown that the approximations
of \sectionname\sectionname \ref{sec:vbc} and \ref{sec:lumfunc}
adequately describe the formation and baryon fraction of the first
galaxies \citep{Naoz09a,Oleary12a}, potentially even underestimating
the suppression in the star-forming gas fraction at the high-mass end
($\approx 10^7 M_\odot$; \citealt{2012arXiv1204.1345M}).

To determine whether the effect of the supersonic baryon--CDM flow on
the abundance of luminous low-mass galaxies persists to the satellite
luminosity function of a Milky-Way sized halo today, we have run
extended Press-Schechter simulations of the merger history of a
10$^{12} M_\odot$ halo, taking into account the effect of the
supersonic flow on the halo mass function and star formation prior to
reionization. We found that the effect largely remains the same and
that the number of satellite galaxies with $M_V \gtrsim -4$ in a
Milky-Way sized halo is typically suppressed by 50\,percent. When then
predicting the observed number of faint satellites of the Milky Way
that could have been observed by the SDSS---multiplying the predicted
counts by the fraction of the effective volume of the Milky Way
observed by the SDSS---we find that a typical initial relative
velocity of $\approx 30$ km s$^{-1}$ alleviates most of the
discrepancy at the faint end ($M_V > -6$) between the observed number
of satellites and that predicted by a model without pre-reionization
suppression of star formation, without applying any additional
mass-dependent reduction in star-formation efficiency, e.g., due to
the radiative feedback from the first stars. Therefore, the effect of
the supersonic baryon--CDM flow naturally provides the amount and
mass-dependence of the pre-reionization suppression of star-formation
efficiency needed to explain the observed luminosity function of
Milky-Way satellites. The limited detection efficiency of the SDSS for
low-luminosity galaxies is such that satellites at $M_V = -2$ can only
be detected out to approximately 40 kpc, such that a large fraction of
the virial volume of the Milky Way remains unexplored at these
low-luminosities. Next-generation surveys such as the LSST will be
able to observe $M_V = 0$ satellite galaxies out the virial radius of
the Milky Way \citep{2008ApJ...688..277T}, and the effect of the
supersonic baryon--CDM flow should lead to a clear suppression in the
number of low-luminosity satellites.

{\bf Acknowledgments:} It is a pleasure to thank Simone Ferraro, Wayne
Hu, and Matias Zaldarriaga for useful comments. Support for Program
number HST-HF-51285.01-A was provided by NASA through a Hubble
Fellowship grant from the Space Telescope Science Institute, which is
operated by the Association of Universities for Research in Astronomy,
Incorporated, under NASA contract NAS5-26555. CD was supported by the
National Science Foundation (NSF) Grant No. AST-0807444, NSF Grant
No. PHY-0855425, and the Raymond and Beverly Sackler Funds.

\bibliographystyle{apj}
\bibliography{vbcbib}

\end{document}